\documentclass[10pt, journal]{IEEEtran}
\usepackage{cite}
\usepackage{bm}
\usepackage{amsmath}
\usepackage{extarrows}
\usepackage{amssymb}
\usepackage{graphicx}
\usepackage{color}
\usepackage{enumerate}
\usepackage{bookmark} 
\usepackage{booktabs}
\graphicspath{{figure}}
\usepackage{setspace}
\usepackage{subfigure}
\usepackage{algorithm}
\usepackage{algpseudocode}
\usepackage{algorithmicx}
\usepackage{multirow}
\usepackage{stfloats}
\usepackage{graphbox}
\usepackage{bbm}
\usepackage{framed} 

\newcommand{\BE}{\begin{equation}}
\newcommand{\EE}{\end{equation}}
\newcommand{\BS}{\begin{subequations}}
\newcommand{\ES}{\end{subequations}}
\renewcommand{\bf}{\bm}

\newtheorem{proposition}{Proposition}
\newtheorem{assumption}{Assumption}

\newtheorem{lemma}{Lemma}

\allowdisplaybreaks \allowdisplaybreaks[2]

\ifCLASSINFOpdf
\graphicspath{{figure/}}

\else

\fi
\begin{document}

\title{Distributed Memory Approximate Message Passing}

\author{\IEEEauthorblockN{Jun Lu, Lei Liu, \emph{Senior Member, IEEE}, Shunqi Huang, \emph{Graduate Student Member, IEEE},  \\Ning Wei, and Xiaoming Chen, \emph{Senior Member, IEEE}}

\thanks{Jun Lu, Lei Liu, and Xiaoming Chen are with the Zhejiang Provincial Key Laboratory of Information Processing, Communication and Networking, College of Information Science and Electronic Engineering, Zhejiang University, Hangzhou, 310007, China (e-mail: \{lu\_jun, lei\_liu, chen\_xiaoming\}@zju.edu.cn).}
\thanks{Shunqi Huang is with the School of Information Science, Japan Advanced Institute of Science and Technology (JAIST), Nomi, Ishikawa 923-1292, Japan (e-mail: shunqi.huang@jaist.ac.jp).}
\thanks{Ning Wei is with the ZTE Corporation and  State Key Laboratory of Mobile Network and Mobile Multimedia Technology, Shenzhen, P.R., 518055, China (e-mail: wei.ning@zte.com.cn).}}

\maketitle
\begin{abstract}
Approximate message passing (AMP) algorithms are  iterative methods for signal recovery in noisy linear systems. In some scenarios, AMP algorithms need to operate within a distributed network. To address this challenge, the distributed extensions of AMP (D-AMP, FD-AMP) and orthogonal/vector AMP (D-OAMP/D-VAMP) were proposed, but they still inherit the limitations of centralized algorithms. In this letter, we propose distributed memory AMP (D-MAMP) to overcome the IID matrix limitation of D-AMP/FD-AMP, as well as the high complexity and heavy communication cost of D-OAMP/D-VAMP. We introduce a matrix-by-vector variant of MAMP tailored for distributed computing. Leveraging this variant, D-MAMP enables each node to execute computations utilizing locally available observation vectors and transform matrices. Meanwhile, global summations of locally updated results are conducted through message interaction among nodes. For acyclic graphs, D-MAMP converges to the same mean square error performance as the centralized MAMP. 
\end{abstract}


\section{Introduction}
\IEEEPARstart{R}{ecovering} a signal from a noisy linear model is a fundamental problem with wide-ranging applications \cite{2021MP_OTFS}. The model can be expressed as:
\BE\label{Eqn:linear}
\bf{y} = \bf{A}\bf{x} + \bf{n},
\EE
where $\bf{y}\!\in\!\mathbb{C}^{M\!\times\!1}$ is the known observation vector, $\bf{A}\!\in\!\mathbb{C}^{M\!\times\! N}$ is a known transform matrix, $\bf{x}$ is the signal to be recovered and $\bf{n}\!\sim\!\mathcal{CN}(\mathbf{0},\sigma^2\bf{I}_M)$ is a Gaussian noise.\par

Approximate message passing (AMP) \cite{donoho2009message, bayati2011dynamics} is an iterative method for solving the problem in \eqref{Eqn:linear}. It is Bayes-optimal for IID Gaussian matrices with a unique state evolution (SE) fixed point (This assumption is maintained throughout) but performs poorly or even diverges for matrices that have correlated entries. Orthogonal/vector AMP (OAMP/VAMP) \cite{ma2017orthogonal, rangan2019vector} overcomes the limitation of AMP and achieves replica Bayes-optimal \cite{takeuchi2020rigorous} performance for right-unitarily-invariant matrices. However, the linear minimum mean square error (LMMSE) estimator in OAMP/VAMP requires a matrix inversion, resulting in high complexity ${\cal O}(M^3+M^2N)$. Convolutional AMP \cite{takeuchi2020convolutional} offers low complexity $\mathcal{O}(MN)$ as AMP but fails to converge for matrices with large condition numbers. Memory AMP (MAMP) \cite{liu2022memory} addresses these weaknesses, achieving replica Bayes-optimal performance for right-unitarily-invariant matrices and guaranteed SE convergence with $\mathcal{O}(MN)$ complexity per iteration.

To address high-dimensional signal processing challenges, distributed AMPs have gained attention. 
Distributed AMP (D-AMP) \cite{DAMP} extends centralized AMP to a centralized distributed network with a central node  for global summation. Fully distributed AMP (FD-AMP) \cite{FDAMP} utilizes summation propagation, derived from consensus propagation \cite{ConsensusPropagation}, to achieve global summation for decentralized distributed networks. Decentralized generalized AMP (D-GAMP) \cite{D-GAMP} has recently been proposed for generalized linear models. However, these distributed algorithms require the transform matrices to be IID Gaussian. Distributed OAMP/VAMP (D-OAMP/D-VAMP) \cite{DOAMP}, \cite{D-VAMP} was proposed to overcome this limitation. However, D-OAMP/D-VAMP inevitably requires matrix inversions and the entire transform matrix for the LMMSE estimator. Thus, the need for nodes to exchange local transform matrices results in considerable communication overhead and storage burden.\par

In this letter, we propose distributed MAMP (D-MAMP) tailored for distributed scenarios. D-MAMP eliminates the need for distributed nodes to interact with local observation vectors and transform matrices, thereby significantly reducing communication overhead and storage burden. We convert the centralized MAMP into a single-step matrix-by-vector form, enhancing its applicability for distributed computing. In D-MAMP, matrix-by-vector computations are performed locally at distributed nodes, leveraging local observation vectors and transform matrices. Global operations, such as summations and combinations of locally updated vector messages, are conducted through message interaction among nodes. The proposed D-MAMP is shown to be replica Bayes-optimal as it converges to the centralized MAMP. Simulation results validate its effectiveness.
\section{Preliminaries}
\subsection{System Model}
\begin{figure}[t]
\centering 
\subfigure[]{
\includegraphics[width=0.44\columnwidth]{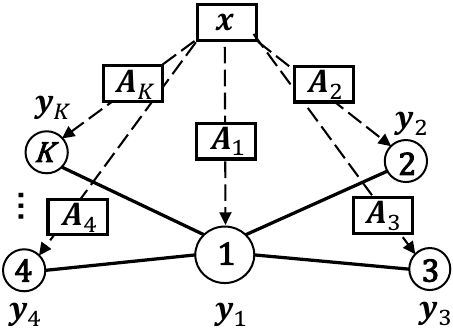}}\vspace{-1mm}\subfigure[]{\includegraphics[width=0.46\columnwidth]{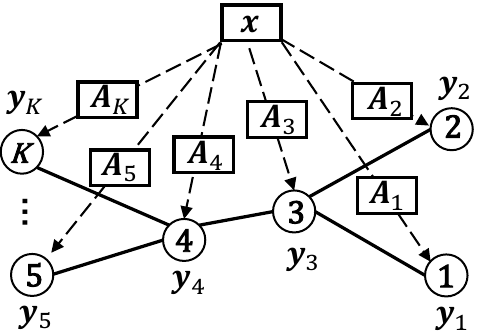}}
\vspace{-2mm}
\caption{(a) Centralized distributed network, (b) Decentralized distributed network.}
\label{Fig:scenarios}
\end{figure}

This letter explores two loopless distributed scenarios: Fig. \ref{Fig:scenarios}(a) shows a centralized distribution network with a central node connected solely to child nodes; Fig. \ref{Fig:scenarios}(b) depicts a decentralized distribution network where each node connects to its adjacent nodes. The network's diameter is defined as the maximum of the minimum distances between any two nodes. A network consists of $K$ nodes, and node $k$ observes $\bf{y}_k = \bf{A}_k\bf{x} + \bf{n}_k$, where {$\bf{A}_k\!\in\!\mathbb{C}^{M_k\!\times\!N}$} are the transform matrices, {$\bf{n}_k\!\in\!\mathbb{C}^{M_k\!\times\!1}$} are the noise vectors, and $\sum_{k=1}^{K}M_{k} = M$. 
\begin{assumption}\label{A1}
We consider large systems with $M,N\to\infty$ and a fixed compression ratio $\delta=M/N$. $\bf{A}$ is a known right-unitarily-invariant matrix\footnote{Let the singular value decomposition of $\bf{A}$ be $\bf{A}=\bf{U\Sigma V}^\mathrm{H}$. Then $\bf{V}$ is Haar distributed and independent of $\bf{U\Sigma}$.}. The entries of $\bm{x}$ are IID with zero mean and normalized variance, i.e., $\tfrac{1}{N}{\rm E}\{\|\bm{x}\|^2\}=1$, and the $(2+i)$-th moments of $\bm{x}$ are finite for some $i > 0$.
\end{assumption}
\subsection{Overview of MAMP}
Before delving into D-MAMP, we provide an overview of the key aspects of centralized MAMP. Let $\lambda_{\mathrm{max}}$ and $\lambda_{\mathrm{max}}$ denote the minimal and maximal eigenvalues of $\bf{AA}^\mathrm{H}$, respectively, and $\lambda_t = \frac{1}{N}\mathrm{tr}\{(\bf{AA}^\mathrm{H})^t\}$. Both of these can be approximately obtained with a low complexity $\mathcal{O}(MN)$ \cite{liu2022memory}.\par
\begin{framed}
\emph{\textbf{MAMP Algorithm}}: Let $\lambda^{\dagger}=(\lambda_{\mathrm{max}}+\lambda_{\mathrm{min}})/2$, $\bf{B}=\lambda^\dagger \bf{I}-\bf{AA}^\mathrm{H}$, and
\vspace{-2mm}
\BS\label{Eqn:MAMP_Org}\BE\label{Eqn:MAMP_LE}
 {\bf{z}}_{t} = 
 \theta_t\bf{B}{\bf{z}_{t-1}} + \xi_t(\bf{y} - \bf{A}\bf{x}_t).
\EE
Starting with $\bf{x}_1=\bf{0}$ and ${\bf{z}_0}=\bf{0}$,
\BE\label{Eqn:MAMP_MLE}
\mathrm{MLE:}\;\bf{r}_t=\frac{1}{\varepsilon_t}\Big(\bf{A}^\mathrm{H}{\bf{z}_t}+\textstyle\sum_{i=1}^tp_{t,i}\bf{x}_i\Big),
\EE
\BE\label{Eqn:MAMP_NLE}
\!\!\!\mathrm{NLE:}\;
\bf{x}_{t+1}=\bar{\phi}_t(\bf{r}_t)=[\bf{x}_1, \cdots ,\bf{x}_t,\phi_t(\bf{r}_t)]\cdot\bf{\zeta}_{t+1},
\EE\ES
where $\phi\left(\cdot\right)$ is the same as that in OAMP/VAMP \cite{ma2017orthogonal}.
\end{framed}
The specific parameters are computed as follows:
\subsubsection{Relaxation}
\BE
    \theta_t =(\lambda^\dagger+\sigma^2/v_{t,t}^{\bar{\phi}})^{-1}  \quad {\rm and} \quad
    \xi_t =\frac{c_{t,2}c_{t,0}+c_{t,3}}{c_{t,1}c_{t,0}+c_{t,2}},
\EE
where $v_{t,t}^{\bar{\phi}}$ is given in \eqref{Eqn.varance_x}, and $c_{t,0}, \cdots, c_{t,3}$ are given in \cite{liu2022memory}.
\subsubsection{Orthogonalization}
\BS\BE
    p_{t,i}=\vartheta_{t,i}w_{t-i}   \quad {\rm and} \quad
    \varepsilon_{t} =\textstyle\sum_{i=1}^{t}p_{t,i},
\EE
where
\vspace{-2mm}
\begin{align}
    b_{i}&\equiv\tfrac{1}{N}\mathrm{tr}\{\bf{B}^{i}\}=\textstyle\sum_{i=0}^{t}\binom{t}{i}(-1)^{i}(\lambda^{\dagger})^{t-i}\lambda_{i},\\
    w_{i}&\equiv\tfrac{1}{N}\mathrm{tr}\{\bf{A}^{\mathrm{H}}\bf{B}^{i}\bf{A}\}=\lambda^{\dagger}b_{i}-b_{i+1},\\ \vartheta_{t,i}&\equiv\xi_i\textstyle\prod_{\tau=i+1}^t\theta_\tau.
\end{align}\ES
\subsubsection{Damping}
For $1\leq t'< t+1$,
\BS\begin{align}
    &v_{t+1,t+1}^\phi\overset{\rm a.s.}{=}[\tfrac{1}{N}\bar{\bf{h}}_{t+1}^\mathrm{H}\bar{\bf{h}}_{t+1}-\delta\sigma^2]/w_0, \\
     \label{Eqn:variance_hpi}
    &v_{t+1,t'}^\phi=(v_{t',t+1}^\phi)^*\overset{\rm a.s.}{=}[\tfrac{1}{N}\bar{\bf{h}}_{t+1}^\mathrm{H}\bf{h}_{t'}-\delta\sigma^2]/w_0, 
\end{align}\ES
where $\bar{\bf{h}}_{t+1}=\bf{y}-\bf{A}\phi_t(\bf{r}_t)$, $\bf{h}_{t'}=\bf{y}-\bf{A}\bf{x}_{t'}$. Suppose that $\bf{V}_{t+1}^\phi$ is invertible\footnote{If $\bm{V}_{t+1}^{\phi}$ is singular, the back-off damping is employed \cite{liu2022memory}.}, we have
\vspace{-3mm}
\BE
    \bf{\zeta}_{t+1}=\frac{(\bf{V}_{t+1}^\phi)^{-1}\bf{1}}{\bf{1}^\mathrm{T}(\bf{V}_{t+1}^\phi)^{-1}\bf{1}}.
\EE
 Let $\bf{v}_{t+1}^{\bar{\phi}}=[v_{t+1,1}^{\bar{\phi}}\cdots v_{t+1,t+1}^{\bar{\phi}}]^{\mathrm{T}}$, we have
\vspace{-2mm}
\BE\label{Eqn.varance_x}
    \bf{v}_{t+1}^{\bar{\phi}}=\frac{\bf{1}}{\bf{1}^\mathrm{T}(\bf{V}_{t+1}^\phi)^{-1}\bf{1}}.
\EE\par
 
\begin{lemma}[Convergence and Bayes Optimality \cite{liu2022memory}]\label{Lemma}
Suppose that Assumption \ref{A1} holds. MAMP converges to the same fixed point as OAMP/VAMP, i.e., it is replica Bayes optimal if it has a unique SE fixed point.
\end{lemma} 
\vspace{-2mm}
\subsection{Consensus propagation}\label{Sec:Con_Prop}

Consider a  loopless bidirectional graph with $K$ nodes, where each node has an initial state value $x_k\in\!\mathbb{R}$. Consensus propagation is used to achieve an average consensus $\frac{1}{K}\sum_{k=1}^{K}x_{k}$. The updates of variables at the $t$-th iteration are as follows:
\vspace{-1mm}
\BS\begin{align}
  {\omega_{k\to j}^{[t]}}&=\frac{x_k+\sum_{i\in\mathcal{N}_k\setminus j}\upsilon_{i\to k}^{[t-1]}\omega_{i\to k}^{[t-1]}}{1+\sum_{i\in\mathcal{N}_k\setminus j}\upsilon_{i\to k}^{[t-1]}},\\ \upsilon_{k\to j}^{[t]}&=1+\sum_{i\in\mathcal{N}_k\setminus j}\upsilon_{i\to k}^{[t-1]},\\ \vspace{-1mm}
{\hat{\omega}_{k}^{[t]}}&=\frac{x_k+\sum_{i\in\mathcal{N}_k\setminus j}\upsilon_{i\to k}^{[t]}\omega_{i\to k}^{[t]}}{1+\sum_{i\in\mathcal{N}_k\setminus j}\upsilon_{i\to k}^{[t]}},
\end{align}\ES
where ${\omega_{k\to j}^{[t]}}$ denotes the state value from node $k$ to node $j$, $\upsilon_{k\to j}^{[t]}$ represents the number of the messages ${\omega_{k\to j}^{[t]}}$, and $\mathcal{N}_k$ denotes the adjacent nodes of node $k$. When the number of iterations equals the graph's diameter $D$, the estimated consensus $\hat{\omega}_{k}^{[D]}$ converges to the exact global average.

\section{Distributed Memory Approximate Message Passing (D-MAMP)}
In this section, we propose D-MAMP for reconstructing $\bf{x}$ from local observations $\bf{y}_k$ in a distributed manner. We present a matrix-by-vector variant of MAMP designed for distributed computing and elaborate on its extension to distributed scenarios. Additionally, summation computation approaches are explored for setups both with and without a central node.\par
\vspace{-2mm}
\subsection{Variational MAMP}
\emph{Challenge:} The original MAMP in \eqref{Eqn:MAMP_Org} relies solely on matrix-vector multiplications, making it well-suited for the distributed computation. However, the MLE in \eqref{Eqn:MAMP_LE} requires knowledge of $\bf{A}$ to compute $\bf{AA}^\mathrm{H}\bf{z}_t$. This imposes significant communication overhead as every node needs access to $\bf{A}$, thereby increasing interaction and associated costs.\par

\emph{Variational MAMP:} Introducing intermediate vectors $\hat{\bf{r}}_t=\bf{A}^{\rm H}{{\bf z}_t}$ enables us to decompose the computation of $\bf{AA}^\mathrm{H}\bf{z}_t$ into two single-step matrix-by-vector operations. In addition, compared to the MLE in the original MAMP, the proposed variational MLE requires one less matrix-by-vector operation.
\begin{framed}
\emph{\textbf{Variational MAMP}}: Starting with $\bf{x}_1=\hat{\bf{r}}_0=\bm{0}$ and $ \hat{\bf{z}}_0=\bm{0}$,
\vspace{-3mm}
\BS\begin{align}\label{Eqn:RMAMP-LE1}
    \!\!\! {\rm MLE:}\; &\hat{\bf{z}}_t=\theta_t\lambda^\dagger\hat{\bf{z}}_{t-1}+\xi_t\bf{y}-\bf{A}(\theta_t\hat{\bf{r}}_{t-1}+\xi_t\bf{x}_t),\\ \label{Eqn:RMAMP-LE2}
    &\hat{\bf{r}}_t=\bf{A}^{\rm H}{ \hat{\bf z}_t},\\ \label{Eqn:RMAMP-LE3}
    &\bf{r}_t=\frac{1}{\varepsilon_t}\left(\hat{\bf{r}}_t + \textstyle\sum_{i=1}^tp_{t,i}\bf x_i\right),\\
    \!\!\! {\rm NLE:}\; &\bf{x}_{t+1} =[\bf{x}_1, \cdots ,\bf{x}_t,\phi_t(\bf{r}_t)]\cdot\bf {\zeta}_{t+1}.\label{Eqn:RMAMP-NLE} 
\end{align}\ES
\end{framed}
\vspace{-3mm}
\subsection{Distributed MAMP (D-MAMP)}
In this subsection, we extend the variational MAMP to D-MAMP, which includes local computation at each node and global summation between nodes.\par
$\bullet$ \emph{Local Computation in MLE}: Node $k$ updates $\hat{\bm{r}}_k$ using its local observation $\bm{y}_k$ and local transform matrix $\bm{A}_k$:
\vspace{-1mm}
\BS\label{Eqn:RDMAMP}
\begin{align}
\hat{\bm{z}}_{t,k}=\theta_t\lambda^+\hat{\bm{z}}_{t-1,k}+&\xi_t\bf{y}_k-\bm{A}_k(\theta_t\hat{\bm{r}}_{t-1}+\xi_t\bf{x}_t),
\\
\hat{\bm{r}}_{t,k}&=\bm{A}_k^{\mathrm{H}}\hat{\bf{z}}_{t,k}.
\end{align}
\ES \par
$\bullet$ \emph{Global Summation in MLE}: The summation in \eqref{Eqn:RDMAMP-LE3} relies on $\hat{\bm{r}}_k$ from all nodes, requiring  inter-node interact messages. 
\vspace{-1mm}
\BE\label{Eqn:RDMAMP-LE3}
\bf{r}_t=\frac{1}{\varepsilon_{t}}\Big(\textstyle\sum_{k=1}^K\hat{\bm{r}}_{t,k}+\textstyle\sum_{i=1}^{t}p_{t,i}\bm{x}_{i}\Big).
\EE\par
$\bullet$ \emph{Local Computation in NLE}: Each node locally performs NLE for $\bm{x}_{t+1}$:
\vspace{-1mm}
\BE\label{Eqn:RDMAMP-NLE}
\bm{x}_{t+1}=\bar{\phi}_t(\bm{r}_t)=[\bm{x}_1, \cdots ,\bm{x}_t,\phi_t(\bm{r}_t)]\cdot\bm {\zeta}_{t+1}.
\EE
\par
It's clear that the parameters $\{\theta_t, \xi_t, \varepsilon_t, p_{t,i}\}$ can be locally calculated at each node. Nevertheless, calculating the damping vector $\bf{\zeta}_t$ requires the covariance matrix $\bm{V}_{t+1}^{\phi}$, which involves the global $\bf{y}$ and $\bf{A}$. This necessitates distributed computation. \par 
\emph{Distributed Computation of $\bm{V}_{t+1}^{\phi}$}: Define $\hat{\bf{x}}_{t+1}=\phi_t(\bf{r}_t)$. $\bm{V}_{t+1}^{\phi}$ can be computed distributively as follows: \par
\emph{$\bullet$ Local Computation}: Each node uses the local $\bm{y}_k$ and $\bm{A}_k$ to compute the measurement errors $\hat{\bf{h}}_{t,k}$ and $\bm{h}_{t,k}$:
\vspace{-2mm}
\BE
\label{Eqn:variance}
\hat{\bf{h}}_{t,k}=\bf{y}_k-\bf{A}_k\hat{\bf{x}}_{t} \quad {\rm{and}}\quad
\bf{h}_{t,k}=\bf{y}_k-\bf{A}_k\bf{x}_t.
\EE\par
\vspace{-2mm}
\emph{$\bullet$ Global Summation}: The covariances  involve all measurement errors $\{\hat{\bm{h}}_{t,k}, \bm{h}_{t,k},  \forall k\}$ and require global summations of scalar $\{\hat{\bm{h}}_{t+1,k}^\mathrm{H}\hat{\bm{h}}_{t+1,k}, \hat{\bm{h}}_{t+1,k}^\mathrm{H}\bm{h}_{t',k}\}$:
\vspace{-2mm}
\BS\label{Eqn:D-variance}\begin{align}
    v_{t+1,t+1}^\phi&\overset{\rm a.s.}{=}\big[\tfrac{1}{N}\textstyle\sum\limits_{k=1}^K\hat{\bf{h}}_{t+1,k}^\mathrm{H} \hat{\bf{h}}_{t+1,k}-\delta\sigma^2\big]/w_0, \\
    v_{t+1,t'}^\phi&\overset{\rm a.s.}{=}\big[\tfrac{1}{N}\textstyle\sum_{k=1}^K\hat{\bf{h}}^\mathrm{H}_{t+1,k}\bf{h}_{t',k}-\delta\sigma^2\big]/w_0.
\end{align}\ES\par 
Note that the D-MAMP algorithm is mathematically equivalent to the centralized MAMP algorithm in \cite{liu2022memory}, differing only in computational methods. Therefore, we have the following proposition from Lemma \ref{Lemma}.
\begin{proposition}
Suppose that Assumption \ref{A1} holds. D-MAMP \footnote{ The number of consensus propagation per iteration in a decentralized distributed network is greater than or equal to the diameter of the graph.} converges to the same fixed point as centralized MAMP, i.e., it is replica Bayes optimal when it has a unique SE fixed point.
\end{proposition}
\subsubsection{D-MAMP for Centralized Distributed Network}
In the centralized distributed network, global summation is performed at the central node. Each node sends its locally computed values to the central node, which performs the summation and distributes the results back to the nodes. Algorithm 1 summarizes the distributed MAMP (D-MAMP) tailored for the centralized distributed network. \par
\begin{algorithm}[t]
    \caption{D-MAMP for centralized distributed network}
    \begin{algorithmic}[1]\label{Alg:MAMP}\small
    \renewcommand{\algorithmicrequire}{\textbf{Input:}}
    \renewcommand{\algorithmicensure}{\textbf{Output:}}
    \Require $\bm{A}_k$, $\bm{y}_k$, $P_x$, $\{\lambda_{\min}, \lambda_{\max}, \lambda_t\}$, $\sigma^{2}$, ${T}$, $L$, $K$. \\\vspace{1mm}
    \hspace{-3mm}\textbf{Initialization:} ${\bm{h}}_{1,k} \!=\!\bm{y}_k,\; \hat{\bm{r}}_0 \!=\!\bm{0}$, $\hat{\bm{z}}_0 =\bm{0}$. 
   \For {$t =1$ to $T$ } \\\% Memory \;LE
   \State update $\hat{\bm{z}}_{t,k}$ and $\hat{\bm{r}}_{t,k}$ (k=1,...,K) according to \eqref{Eqn:RDMAMP}.
  \State $\hat{\bm{r}}_t=\sum_{k=1}^{K}\hat{\bm{r}}_{t,k}
 \hspace{2cm} \%\textbf{Global Computation}$
  \State update $\bm{r}_t$ according to \eqref{Eqn:RDMAMP-LE3}.\\\% NLE 
  \State  $(\hat{\bm{x}}_{t+1}, \hat{v}_{t+1})= \phi_t(\bm{r}_t)$   
  \State update $\hat{\bm{h}}_{t+1,k}$ and $\bf{h}_{t,k}$ according to \eqref{Eqn:variance}.
  \State update$\bm{V}_{t+1}^{\phi}$ according to \eqref{Eqn:D-variance}. \%\textbf{Global Computation}
  \State update $\bm{x}_{t+1}$ according to \eqref{Eqn:RDMAMP-NLE}. 
    \EndFor
    \vspace{1mm}
    \Ensure  $\{\hat{\bm{x}}_{t+1}, \hat{v}_{t+1} \}$.
 \end{algorithmic}
\end{algorithm}
\subsubsection{D-MAMP for Decentralized Distributed Network}
In the decentralized distributed network, global summation is computed utilizing consensus propagation, as detailed in Section \ref{Sec:Con_Prop}. Each node reaches a consensus on the global summation by interacting with adjacent nodes to exchange local computation results. The communication cost is outlined in Table I, where $D'$ represents the number of consensus propagation per iteration. Note that the communication cost of distributed MAMP is solely related to the unknown signal dimension $N$ and is decoupled from the observed signal dimension $M$, which is a crucial consideration for industry applications. Algorithm 2 summarizes fully distributed MAMP (FD-MAMP) tailored for the decentralized distributed network. 
 
\begin{table}[b]\label{Table}
\centering
\vspace{-2mm}
\caption{Properties of distributed MAMP}
\begin{tabular}{|c|c|c|}
\hline  
Algorithms& Complexity& Communication Cost\\
\hline 
D-MAMP& $\mathcal{O}(MN/K)$& $2T(N+1)(K-1)$\\
\hline
FD-MAMP& $\mathcal{O}(MN/K)$& $2TD'(N+1)(K-1)$\\
\hline
\end{tabular}
\end{table}

\begin{algorithm}[t]
\caption{FD-MAMP for decentralized distributed network}
 \begin{algorithmic}[1]\label{Alg:FD-MAMP}\small
 \renewcommand{\algorithmicrequire}{\textbf{Input:}}
 \renewcommand{\algorithmicensure}{\textbf{Output:}}
 \Require $\bm{A}_k$, $\bm{y}_k$, $P_x$, $\{\lambda_{\min}, \lambda_{\max}, \lambda_t\}$, $\sigma^{2}$, ${T}$, $L$, $K$. \\\vspace{1mm}
 \hspace{-3mm}\textbf{Initialization:} ${\bm{h}}_{1,k} \!=\!\bm{y}_k,\; \hat{\bm{r}}_0 \!=\!\bm{0}$, $\hat{\bm{z}}_0 =\bm{0}$.
   \For {$t =1$ to ${T}$ } \\ \% Memory \;LE
  \State update $\hat{\bm{z}}_{t,k}$ and $\hat{\bm{r}}_{t,k}$ (k=1,...,K) according to \eqref{Eqn:RDMAMP}.
  \State update $\hat{\bm{r}}_t$ by \textbf{Consensus Propagation}
  \State update $\bm{r}_t$ according to \eqref{Eqn:RDMAMP-LE3}.\\\% NLE 
\State  $(\hat{\bm{x}}_{t+1}, \hat{v}_{t+1})= \phi_t(\bm{r}_t)$  
  \State update $\hat{\bm{h}}_{t+1,k}$ and $\bf{h}_{t,k}$ according to \eqref{Eqn:variance}.
  \State update$\bm{V}_{t+1}^{\phi}$ by \textbf{Consensus Propagation}
  \State update $\bm{x}_{t+1}$ according to \eqref{Eqn:RDMAMP-NLE}.
  \EndFor \vspace{1mm} 
 \Ensure $\{\hat{\bm{x}}_{t+1}, \hat{v}_{t+1} \}$.
 \end{algorithmic}
\end{algorithm} 
\vspace{-3mm}
\subsection{Distributed Computation of \texorpdfstring{$\{\lambda_t\}$}{}} \label{Sec:App_lamd}
MAMP requires parameters $\{\lambda_{\min}, \lambda_{\max}\}$ and $\lambda_t$, where $\lambda_t$ can be efficiently approximated using a recursive method. $\lambda_{\min}$ and $\lambda_{\max}$ can be replaced by a lower bound $\lambda_{\min}^{\rm low}=0$ and an upper bound $\lambda_{\max}^{\mathrm{up}}=(N\lambda_{\tau})^{1/\tau}$\cite{liu2022memory}, which becomes tighter for larger $\tau$. Typically, $\tau$ is set to twice the maximum number of iterations $T$. However, computing $\lambda_t$ relies on the global transform matrix $\bf{A}$, which is not accessible at distributed nodes. To address this, we propose an efficient  distributed approximation method of $\lambda_t$ in Proposition \ref{Pro:2}. 
\begin{proposition}\label{Pro:2}
We approximate  $\lambda_t$ by 
\vspace{-1mm}
\BE
    \lambda_t\stackrel{\mathrm{a.s.}}{=}\lim_{N\to\infty}\|\bf{s}_t\|^2,
\EE
where $\bf{s}_t$ is obtained by the recursion: Starting with $t=1$ and $\bf{s}_{0}\sim\mathcal{CN}(\mathbf{0}_{N\times1},\frac{1}{N}\mathbf{I}_{N\times N})$, 
\vspace{-1mm}
\BS\begin{align}\label{Eqn:D-approximate-1}
 {\rm Local:}\;\; {\bf{s}}_{t,k}&=
\begin{cases}
\bf{A}_{k}{\bf{s}}_{t-1}, &\; {\rm if}\; t {\rm \;is\; odd} \vspace{1.5mm}\\
\bf{A}_{k}^{\mathrm{H}}{\boldsymbol{s}}_{t-1}, & \; {\rm if}\; t {\rm \;is\; even},
\end{cases}\\
{\rm Global:}\;\;\; {\bf{s}}_{t}&=
\begin{cases}
[{\bf{s}}^{\mathrm{T}}_{t,1},\dots, {\bf{s}}^{\mathrm{T}}_{t,K} ]^{\mathrm{T}}, &\; {\rm if}\; t {\rm \;is\; odd} \vspace{1.5mm}\\
\sum_{k=1}^{K} {\bm{s}}_{t,k}, & \; {\rm if}\; t {\rm \;is\; even},
\end{cases}\label{Eqn:D-approximate-sum1}\end{align}
\ES
where for odd $t$, ${\bf{s}}_t\in\mathbb{C}^{M}$ and ${\bf{s}}_{t,k}\in\mathbb{C}^{\frac MK}$, while for even $t$, ${\bf{s}}_{t}\in\mathbb{C}^N$ and ${\bf{s}}_{t,k}\in\mathbb{C}^{N}$.
\end{proposition}\par 
\emph{proof:} Retracing the above recursive process, $\bf{s}_t$ can be expressed as
\vspace{-3mm}
\BE
{\bf{s}}_t=
\begin{cases}
\bf{A}(\bf{A}^{\mathrm{H}}\bf{A})^{\frac{t-1}{2}}{\bf{s}}_{0}, &\; {\rm if}\; t {\rm \;=\;1, 3, 5,\cdots} \vspace{1.5mm}\\
(\bf{A}^{\mathrm{H}}\bf{A})^{\frac{t}{2}}{\bf{s}}_{0}, &\; {\rm if}\; t {\rm \;=\;2, 4, 6,\cdots}.
\end{cases}
\EE
We have
\vspace{-2mm}
\BS\begin{align}
    \lim_{N\to\infty}\|\bf{s}_{t}\|^{2}\stackrel{a.s.}{=}&\lim_{N\to\infty}\bf{s}_{0}^{\mathrm{H}}(\bf{A}^{\mathrm{H}}\bf{A})^{t}\bf{s}_{0}\\=&\frac{1}{N}{\rm tr}\{(\bf{A}^{\mathrm{H}}\bf{A})^{t}\}=\lambda_{t}.
\end{align}\ES
Thus, we have completed the proof. \par

In distributed scenarios, each node computes \eqref{Eqn:D-approximate-1} locally and then interacts its results with others for global summation and combination as in \eqref{Eqn:D-approximate-sum1}. For centralized and decentralized distributed networks, the distributed approximation of $\lambda_t$  incurs communication cost as follows:
\BS\begin{align}
&{\rm Centralized:}\;\;\;\;\;  T(K-1)(2N+M+M/K),\\
&{\rm Decentralized:}\;\;   2TD'(K-1)(M+N). 
\end{align}
\ES
\begin{figure*}[t]
	\centering
        \begin{minipage}{0.32\linewidth}
        \centering
        \vspace{-3mm}
	\includegraphics[width=1\columnwidth]{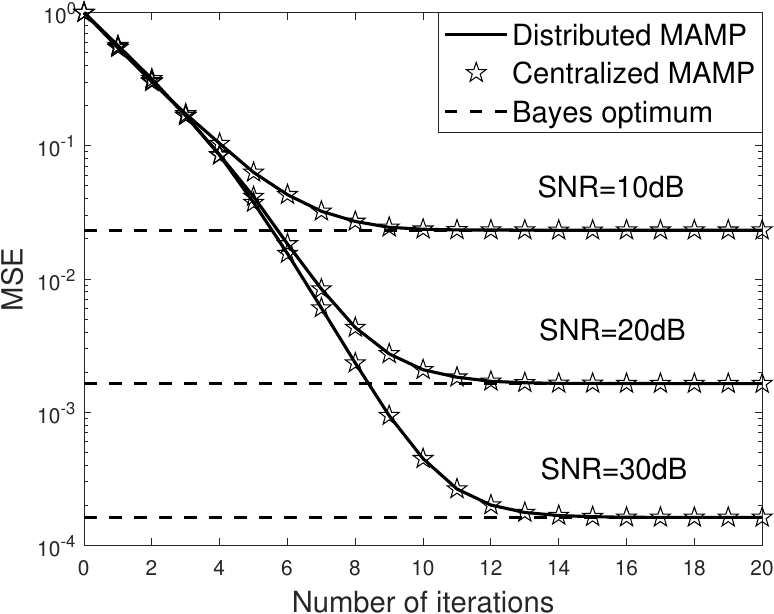} 
    \vspace{-6mm}
    \caption{MSE versus the number of iterations for centralized MAMP and D-MAMP. $M_k=125$, $N=2000$, $\kappa=10$.}\label{Fig:DMAMP}
        \end{minipage}\quad
        \begin{minipage}{0.32\linewidth}
	\centering
 \vspace{-3mm}
    \includegraphics[width=1\columnwidth]{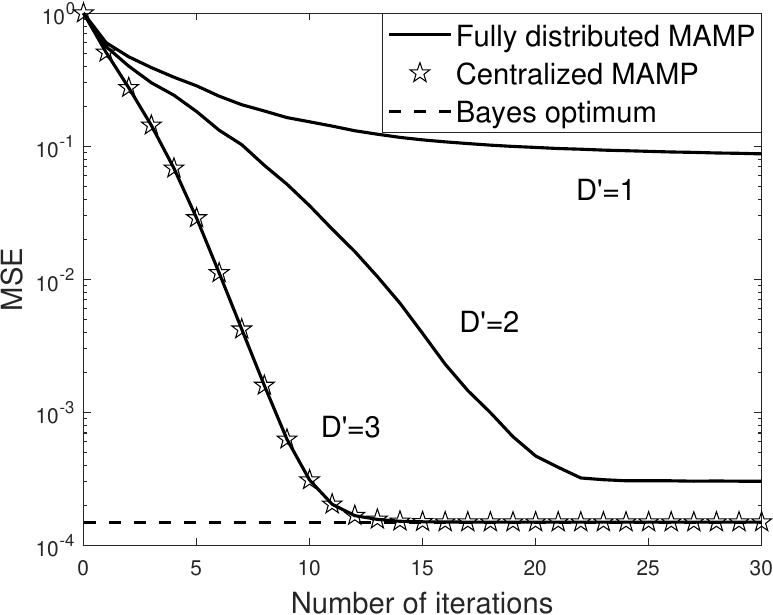}  
    \vspace{-6mm}
    \caption{MSE versus the number of iterations for centralized MAMP and FD-MAMP. $M_k=125$, $N=2000$, $\kappa=10$, $\rm{SNR = 30dB}$.}\label{Fig:FD-MAMP}
       \end{minipage}\quad
       \begin{minipage}{0.32\linewidth}
	\centering
\vspace{-3mm}	\includegraphics[width=1\columnwidth]{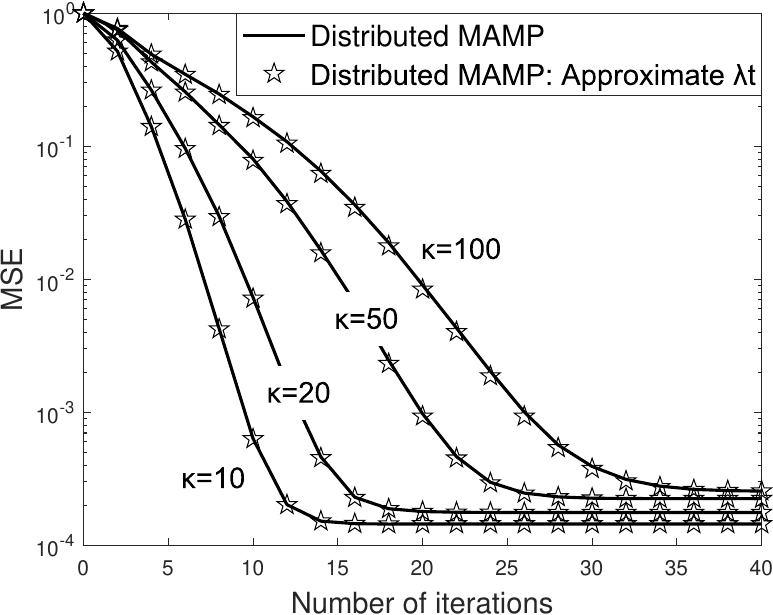}
    \vspace{-6mm}
    \caption{MSE versus the number of iterations for D-MAMP with exact $\{\lambda_{\min}, \lambda_{\max}\}$ and approximate $\lambda_{t}$. $M_k=125$, $N=2000$, $\rm{SNR = 30dB}$.}\label{Fig:App_lamda_kappa.png}
    \end{minipage}   
\end{figure*}
\vspace{-1cm}
\section{simulation Results}\label{sec:num}
In this section, we assess the performance of the D-MAMP and decentralized MAMP via computer simulations. This simulation constructs a distributed compressed sensing problem with sparse Bernoulli-Gaussian signals, where each element ${x}_i$ of $\bf{x}$ follows $p(x_i)=\mu\cdot G(x_i;\mu^{-1})+(1-\mu)\cdot\delta(x_i)$. The ${G(x;v)}$ and $\delta(x)$ denote the zero-mean Gaussian distribution with variance $v$ and Dirac delta function, respectively. The parameter $\mu\in[0,1]$ determines the sparsity of the signal $\bm{x}$. We adhere to the approach outlined in centralized MAMP \cite{liu2022memory} to obtain the right-unitarily-invariant matrix $\bm{A}$ with the condition number $\kappa$. We fix the number of nodes to $K=8$ and specifically configured the diameter of the decentralized distributed network graph to $D = 3$.\par

Fig. \ref{Fig:DMAMP} illustrates MSE versus the number of iterations for centralized MAMP and D-MAMP at various SNRs. As evident, D-MAMP can synchronize with centralized MAMP to converge to the same point without any performance loss.\par
In Fig. \ref{Fig:FD-MAMP}, we depict MSE versus the number of iterations for FD-MAMP with varying numbers of consensus propagation per iteration. When $D^{\prime}$ equals the diameter $D$, FD-MAMP exhibits no performance loss compared to centralized MAMP, but when $D^{\prime}$ is smaller, nodes fail to reach consensus on global information, leading to non-convergence of MSE performance degradation.\par

Fig. \ref{Fig:App_lamda_kappa.png} compares the MSE of D-MAMP with exact $\{\lambda_{\min}, \lambda_{\max}\}$ and D-MAMP with approximate
$\lambda_{t}$ in \ref{Sec:App_lamd}. As we can see, the performance of approximation is nearly equivalent to that of knowing the exact values for D-MAMP.\par

\section{Conclusion}
This letter proposes D-MAMP and FD-MAMP for centralized and decentralized distributed networks, respectively. All matrix-by-vector computations are performed locally, while global operations, such as summations and combinations of local messages, are handled through message interaction among the nodes. We demonstrate that D-MAMP and FD-MAMP are replica Bayes-optimal since they are mathematically equivalent to the centralized MAMP. The efficiency of D-MAMP and FD-MAMP is confirmed by simulation results. 

\bibliographystyle{IEEEtran}
\clearpage
\bibliography{reference}

\begin{thebibliography}{10}
\providecommand{\url}[1]{#1}
\csname url@samestyle\endcsname
\providecommand{\newblock}{\relax}
\providecommand{\bibinfo}[2]{#2}
\providecommand{\BIBentrySTDinterwordspacing}{\spaceskip=0pt\relax}
\providecommand{\BIBentryALTinterwordstretchfactor}{4}
\providecommand{\BIBentryALTinterwordspacing}{\spaceskip=\fontdimen2\font plus
\BIBentryALTinterwordstretchfactor\fontdimen3\font minus \fontdimen4\font\relax}
\providecommand{\BIBforeignlanguage}[2]{{%
\expandafter\ifx\csname l@#1\endcsname\relax
\typeout{** WARNING: IEEEtran.bst: No hyphenation pattern has been}%
\typeout{** loaded for the language `#1'. Using the pattern for}%
\typeout{** the default language instead.}%
\else
\language=\csname l@#1\endcsname
\fi
#2}}
\providecommand{\BIBdecl}{\relax}
\BIBdecl

\bibitem{2021MP_OTFS}
Z.~Yuan, F.~Liu, Q.~Guo, and Z.~Wang, ``Message passing based detection for orthogonal time frequency space modulation,'' \emph{ZTE Communications}, vol.~19, no.~4, pp. 34--44, Dec. 2021.

\bibitem{donoho2009message}
D.~L. Donoho, A.~Maleki, and A.~Montanari, ``Message-passing algorithms for compressed sensing,'' \emph{Proc. Natl. Acad. Sci. U.S.A.}, vol. 106, no.~45, pp. 18\,914--18\,919, Nov. 2009.

\bibitem{bayati2011dynamics}
M.~Bayati and A.~Montanari, ``The dynamics of message passing on dense graphs, with applications to compressed sensing,'' \emph{{IEEE} Trans. Inf. Theory}, vol.~57, no.~2, pp. 764--785, Feb. 2011.

\bibitem{ma2017orthogonal}
J.~Ma and L.~Ping, ``Orthogonal {AMP},'' \emph{{IEEE} Access}, vol.~5, pp. 2020--2033, Jan. 2017.

\bibitem{rangan2019vector}
S.~Rangan, P.~Schniter, and A.~K. Fletcher, ``Vector approximate message passing,'' \emph{{IEEE} Trans. Inf. Theory}, vol.~65, no.~10, pp. 6664--6684, May. 2019.

\bibitem{takeuchi2020rigorous}
K.~Takeuchi, ``Rigorous dynamics of expectation-propagation-based signal recovery from unitarily invariant measurements,'' \emph{{IEEE} Trans. Inf. Theory}, vol.~66, no.~1, pp. 368--386, Jan. 2020.

\bibitem{takeuchi2020convolutional}
------, ``Convolutional approximate message-passing,'' \emph{{IEEE} Signal Process. Lett.}, vol.~27, pp. 416--420, Feb. 2020.

\bibitem{liu2022memory}
L.~Liu, S.~Huang, and B.~M. Kurkoski, ``Memory {AMP},'' \emph{{IEEE} Trans. Inf. Theory}, vol.~68, no.~12, pp. 8015--8039, Jun. 2022.

\bibitem{DAMP}
P.~Han, R.~Niu, M.~Ren, and Y.~C. Eldar, ``Distributed approximate message passing for sparse signal recovery,'' in \emph{{IEEE} GlobalSIP}.\hskip 1em plus 0.5em minus 0.4em\relax IEEE, 2014, pp. 497--501.

\bibitem{FDAMP}
R.~Hayakawa, A.~Nakai, and K.~Hayashi, ``Distributed approximate message passing with summation propagation,'' in \emph{Proc. {IEEE} Int. Conf. Acoust., Speech Signal Process. (ICASSP)}.\hskip 1em plus 0.5em minus 0.4em\relax IEEE, 2018, pp. 4104--4108.

\bibitem{ConsensusPropagation}
C.~Moallemi and B.~Van~Roy, ``Consensus propagation,'' \emph{{IEEE} Trans. Inf. Theory}, vol.~52, no.~11, pp. 4753--4766, Nov. 2006.

\bibitem{D-GAMP}
K.~Takeuchi, ``Decentralized generalized approximate message-passing for tree-structured networks,'' in \emph{Proc. {IEEE} Int. Conf. Acoust., Speech Signal Process. (ICASSP)}.\hskip 1em plus 0.5em minus 0.4em\relax IEEE, 2024, pp. 12\,866--12\,870.

\bibitem{DOAMP}
K.~Hisanaga and M.~Isaka, ``Communication-efficient distributed orthogonal approximate message passing for sparse signal recovery,'' \emph{{IEICE} Trans. Fundam. Electron. Comput. Sci.}, vol. 107, no.~3, pp. 493--502, Dec. 2024.

\bibitem{D-VAMP}
M.~Karuppasamy, M.~Akrout, F.~Bellili, and A.~Mezghani, ``Distributed vector approximate message passing,'' in \emph{Proc. {IEEE} Int. Conf. Acoust., Speech Signal Process. (ICASSP)}.\hskip 1em plus 0.5em minus 0.4em\relax IEEE, 2024, pp. 9611--9615.

\end{thebibliography}
\end{document}